\documentclass{elsart}
\usepackage{epsfig}
\usepackage{graphicx}
\usepackage{lscape}
\usepackage{subfigure}
\usepackage{amsmath}

\journal{Physics Letters A}

\begin{document}

\begin{frontmatter}

\title{Anomalies in the Multicritical Behavior of Staggered Magnetic and Direct Magnetic Susceptibilities of Iron Group Dihalides}
\author{Gul Gulpinar  $^{a}$ },
\corauth[cor]{corresponding author.}
\ead{gul.gulpinar@deu.edu.tr}
\author{Erol Vatansever $^{a}$},
\address{Dokuz Eylul University, Department of Physics, 35160-Buca, Izmir, Turkey}

\begin{abstract}
 \vskip0.2cm \hskip1.1cm
 In this study, the temperature dependencies of magnetic response functions of the anhydrous dihalides of iron-group elements
 are examined in the neighborhood of   multi-critical points (tricritical, critical end point, double critical end point)
and first order transition temperatures within molecular field approximation. Our findings reveal the fact that metamagnetic Ising system
exhibits anomalies in the temperature dependence of the magnetic response functions for $r<0,3$.
In addition, we extensively investigated how an inter- and intra-layer exchange interaction ratio can influence  magnetic response properties of these systems.
Finally, a comparison is made with related works.
\end{abstract}


\end{frontmatter}

\section{Introduction}
\indent  Intensive theoretical and experimental effort is devoted to
investigate the multicritical phenomena for more than half a century.
The tricritical point (TCP) is one of the first multicritical points investigated which can be roughly viewed as
a point separating a second order transition line from a  first-order transition line, at which the three
coexisting phases simultaneously become critical. Itinerant ferromagnets \cite{Taufour}, multicomponent fluid mixtures \cite{Hankey}, pentanary microemulsions \cite{Ginzberg}, ammonium chloride \cite{Garland}, and $^{3}He-^{4}He$ mixtures \cite{Riedel}  are other systems that represent tricritical behavior. In addition, it is shown that  there exist tricritical points in an experimentally accessible
three-dimensional space of the electric field, temperature and pressure in ferroelectrics \cite{Peercy}.
On the other hand, a critical end point (CEP) appears when a line of second-order phase transitions terminates
at a first-order phase boundary delimiting a new noncritical phase. At this multicritical point,  a line of second-order
phase transitions intersects and is truncated by a first-order phase boundary, beyond which a new noncritical phase is
formed. Binary alloys \cite{Leidl}, relaxor ferroelectrics \cite{Iwata}, binary fluid mixtures \cite{Konynen}, ferromagnets \cite{Demko}, Ising random-field model \cite{Kauf}, and metamagnets \cite{Cohen,Selke} are the physical systems in which the CEP are common.
In 1997, an extensive Monte Carlo (MC) simulation \cite{Wild}  presented the singular behavior on the first-order transition
line close to CEP in a classical binary fluid \cite{Fisher1ve2,Fisher3,Barbosa}.\\
\indent In addition to TCP and CEP, the double critical (bicritical) end point DCP
appears where two critical lines end simultaneously at a first-order phase boundary. Double critical end points have been
observed in binary and quasi-binary mixtures \cite{Poot}, and there is also some indication of
a DCP in the metamagnet $FeBr_{2}$ \cite{Wolf,Katsumata}. According to mean-field approximation (MFA), the next-nearest-neighbor
Ising antiferromagnetic model, the layered metamagnet and the random-field Ising model present a DCP \cite{Cohen,Kauf,Stryj}.
In addition, The MC simulations exhibited the decomposition of the TCP into a DCP and a CEP  in
three dimensional  spin-$1$ Blume-Capel (BC) model \cite{Wang} while in $d=2$, only a fully stable
TCP is observed \cite{Kimel}. Recently, Plascak and  Landau  studied
a DCP  in the two-dimensional spin-$3/2$ BC model  via extensive
MC simulations \cite{Plascak}. \\
\indent On the other hand, the behavior of the  staggered and direct susceptibilities in the neighborhood of phase transitions has been a subject of experimental and theoretical research for quite a long time: In 1975, a two lattice model of antiferromagnetic phase transitions is discussed in detail, using the Gell-Mann-Low formulation of renormalization group methods and Wilson's $\epsilon$ expansion.  In this study, Alessandrini et. al. have obtained the disordering susceptibility and staggered susceptibility in terms of two-point function at zero magnetic fields and zero momentum \cite{Alessandrini}. Later, Landau has obtained Monte Carlo data for a simple cubic antiferromagnet with nearest- and next-nearest-neighbor interactions which reveal asymptotic tricritical behavior of the order parameter and high-temperature susceptibilities which are mean-field-like without corrections, in agreement with renormalization-group calculations \cite{Landau, Nelson}.
Using the high-temperature series expansion for the extended Hubbard model, Barkowiak et. al. have obtained
the series to the sixth order for the staggered magnetic susceptibility and charge-ordered susceptibility \cite{Barkowiak}.
Recently, Li et.al.  studied the susceptibility of two-dimensional Ising model
on a distorted Kagome lattice  by means of exact solutions and
the tensor renormalization-group (TRG) method \cite{Li}. In addition,
magnetic behaviors of $\beta-Cu_{2}V_{2}O_{7}$ single crystals are
investigated by means of magnetic susceptibility  measurements \cite{He}.
Millis et.al.  report measurements of the magnetization
and susceptibility of a series of samples of two different
variants of the molecular magnet Mn12-ac: the usual, much studied
form referred to as $Mn_{12}$-ac and a new form abbreviated
as $Mn_{12}$-ac-$MeOH$ \cite{Millis}.\\
\indent Metamagnetic materials are of great interest since it is possible to
induce novel kinds of critical behavior by forcing competition between ferromagnetic
and anti-ferromagnetic couplings existing in the system, in particular
by applying a external magnetic field. Magnetic materials that exhibit field-induced transitions
can generally divided in two classes; (i) Highly anisotropic, (ii) Isotropic or weakly anisotropic.
The phase transitions in anisotropic materials (class (i) ) are usually characterized by simple reveals
of the spin directions which are in contrast with transitions in class (ii). The field-induced transitions
in class (ii) materials are related to a rotation of  the local spin directions \cite {Giordano}.
Iron group dihalides; compounds such as $FeCl_{2}$, $FeBr_{2}$, $FeCl_{2}2H_{2}O$, $FeMgBr_{2}$, $CoCl_{2}$ and $NiCl_{2}$ fall in the
first class  \cite{kalita,fujita}. Some theoretical Hamiltonian models describing the behavior of iron group dihalides
have been proposed.  Monte-Carlo simulation \cite{Landau72,Arora}  and high-temperature series expansion calculations
\cite{Harbus1,Harbus2}  have been performed on a simple cubic lattice Ising model with in-plane
ferromagnetic coupling and antiferromagnetic coupling between adjacent planes (the meta model)
and on the next-nearest-neighbor (nnn) model with antiferromagnetic nearest-neighbor (nn) and ferromagnetic next nearest-neighbor (nnn) interactions.
Recently, a MC simulation has been performed on a quite realistic model of $FeCl_{2}$ in a magnetic field
\cite{Hernandez} and this typical metamagnet has also been treated by a high-density expansion method on a two-sublattice
collinear Heisenberg Ising ($s=1$) model with three- and four-ion anisotropy \cite{Onyszkiewicz1,Onyszkiewicz2}. \\
\indent  Although much effort devoted the critical behavior of the metamagnetic systems, to the best of our knowledge, there has been  no studies investigating the temperature and field dependencies of
direct magnetic and staggered magnetic susceptibilities in the neighborhood
of multicritical critical points such as critical end point and
double critical end point as well as first order transition points. \\
\indent The layout of this letter is as follows: The derivation of the expressions describing the  mean field staggered magnetic and magnetic susceptibilities is represented in Section {1}. The results describing the temperature and field dependencies of
the direct and staggered magnetic response functions  are
given Section 3, and finally Section 4 contains the conclusions and discussions.

    \section{Derivation of static staggered magnetic and magnetic susceptibilities of Spin-1/2  Metamagnetic Ising Model}
 \hspace{0.5cm}
 In order to obtain staggered magnetic susceptibility one should introduce a staggered external field $H_{s}$ to the system \cite{Cohen}
 whereas total (direct) magnetic susceptibility is the response of the total magnetization to a physical external field $H$.
Consequently the Hamiltonian of the spin-$\frac{1}{2}$ metamagnetic Ising  Model can be written as
\begin{equation}
\label{1}
  \hat{H}=-\sum_{intra}J_{ij} S_{i}S_{j}-J^{'} \sum_{inter}S_{k}S_{l}-H\sum_{i}S_{i}-H_{s}\sum_{i}(-1)^{i}S_{i} ,
\end{equation}
where the first sum refers to ferromagnetic couplings
$J_{ij}=J>0$ between spins $S_{i}$  and $S_{j}$  in the same x-y layers
and the second sum denotes anti-ferromagnetic
couplings $J^{'}_{kl}=J^{'}<0$ between spins
in  adjacent layers. In addition, $H$ and $H_{s}$ denotes the
external physical magnetic field and external staggered magnetic field respectively.
By making use of mean field approximation  free energy per spin can be obtained as
\begin{equation}
\begin{array}{clc}
 \label{2}
f=&\frac{1}{4}T[(1+m_{a})\ln(1+m_{a})+(1-m_{a})\ln(1-m_{a})\\
\\
&+(1+m_{b})\ln(1+m_{b})+(1-m_{b})\ln(1-m_{b})-4\ln(2)]\\
\\
&+\frac{1}{2}z_{1}J^{'}m_{a}m_{b}-\frac{1}{4}z_{2}J(m_{a}^{2}+m_{b}^{2})-\frac{1}{2}H(m_{a}+m_{b})-\frac{1}{2}H_{s}(m_{a}-m_{b}).
\end{array}
\end{equation}
\indent In the constant magnetic field distribution, the sublattice magnetization
$m_{a}$ and $m_{b}$ are functions of the independent variables $T, H$,and $H_{s}$ so that
free energy per spin represented by Eq.(\ref{2}) is a  non-equilibrium thermodynamic potential which depends on several order variables \cite{lav}.
The equilibrium state corresponds to the minimum of $f$ with respect to $m_{a}$ and $m_{b}$. On the other hand, the conditions for a stationary value of $f$ can  be expressed as follows:
\begin{equation}
\begin{array}{lcl}
\label{3}
\frac{2}{T}\frac{\partial f}{\partial m_{a}}=\kappa(m_{a})+\frac{z_{1}J^{'}}{T}(m_{a}+m_{b})-\frac{z_{1}J^{'}}{T}H -\frac{z_{1}J^{'}}{T}H_{s} =0,\\
\\
\frac{2}{T}\frac{\partial f}{\partial m_{b}}=\kappa(m_{b})+\frac{z_{1}J^{'}}{T}(m_{a}+m_{b})-\frac{z_{1}J^{'}}{T}H+ \frac{z_{1}J^{'}}{T}H_{s}=0.
\end{array}
\end{equation}
and
\begin{equation}
\begin{array}{lcl}
\label{4}
\Delta&=&\frac{4}{T^{2}}\left[\frac{\partial^{2}f}{\partial m_{a}^{2}}\frac{\partial^{2}f}{\partial m_{b}^{2}}-\left(\frac{\partial^{2}f}{\partial m_{a} \partial m_{b}}\right)^{2}\right]\\
\\
&=&\kappa^{'}(m_{a})\kappa^{'}(m_{b})+\tau^{-1}{\kappa^{'}(m_{a})+\kappa^{'}(m_{b})}>0,\\
\end{array}
\end{equation}
where $\kappa(m_{i})$   and  $\kappa^{'}(m_{i})$ are
\begin{equation}
\label{5}
\kappa(m_{i})=\frac{1}{2}\left[\frac{1+m_{i}}{1-m_{i}}\right]-\frac{z_{1}J^{'}}{T}(1+\frac{z_{2}J}{z_{1}J^{'}})m_{i}=-\kappa(-m_{i}),
\end{equation}

\begin{equation}
\label{6}
\kappa^{'}(m_{i})=\frac{1}{1-m_{i}^{2}}-\frac{z_{1}J^{'}}{T}(1+\frac{z_{2}J}{z_{1}J^{'}}).
\end{equation}\\
\indent In order to investigate the behavior of the metamagnetic system in the neighborhood of phase transition points,
it is more convenient to formulate the system in terms of
total and staggered magnetization which are given as follows:
    \begin{equation}  \label{7}
    m_{t}=\frac{m_{a}+m_{b}}{2}, m_{s}=\frac{m_{a}-m_{b}}{2}.
    \end{equation}

 \indent Inserting Eq.(\ref{7}) in Eq.(\ref{3})  one obtains the following mean field equations of state for the spin-1/2  metamagnetic Ising  model on a cubic lattice as below
\begin{equation}\label{7}
\begin{array}{lcl}
\displaystyle m_{t} &=&\displaystyle \frac{\sinh2(\frac{H-m_{t}\delta}{k_{B}T})}{\cosh2(\frac{H-m_{t}\delta}{k_{B}T})+\cosh2(\frac{H_{s}+m_{s}t}{k_{B}T})},
\\
\\
\displaystyle m_{s} &=&\displaystyle \frac{\sinh2(\frac{H_{s}+m_{s}t}{k_{B}T})}{\cosh2(\frac{H-m_{t}\delta}{k_{B}T})+\cosh2(\frac{H_{s}+m_{s}t}{k_{B}T})}.
\end{array}
\end{equation}
 Here, $\delta=J-J'$, $t=J+J'$, $z_{1}=2$ and $z_{2}=4$ .
 The metamagnetic spin-$1/2$ Ising model  exhibits field-induced
 phase transitions. In Fig. 1(a) and (b) the  variation of the antiferromagnetic and ferromagnetic order parameters  are given
 in the field-temperature plane for a second order phase transition for $r=1.0$.

  It has been shown  in the  extensive theoretical review by Kincaid and Cohen that
metamagnetic Ising model exhibits different types of phase boundaries \cite{Cohen}.
In this study,  a Landau expansion of the free energy
is performed and by a careful analysis of the signs of the coefficients, the possibility of different
phase diagrams has been revealed. Further, Moreira et. al. has extended this analysis considering
terms up to twelfth order \cite{Moreira}. The Landau expansion consists in developing the mean-field
free energy given by Eq(\ref{4}) in a power series  of the order parameter ($m_{s}$) which vanishes near the critical point:
\begin{equation}
\Psi(T,H,m_{s})=\sum\limits_{k=0}^{n} \psi_{2k}(T,H) m_{s}^{2k}
\label{10}
\end{equation}
According to the values and signs of the expansion coefficients one can distinguish different
types of phase transitions:
(i) For  $\psi_{2}=0$ and  $\psi_{4}<0$ and $\psi_{6}>0$  a first order transition appears.
(ii) If  $\psi_{2}=0$ and  $\psi_{4}>0$ then an ordinary critical point takes place.
(iii) For $\psi_{2}=0$ and  $\psi_{4}=0$ and $\psi_{6}>0$ we experience tricritical point as a function of the ratio of the exchange interactions ($\eta=\frac{z_{2}J}{z_{1}J^{'}}$):
\begin{equation}\label{11}
  \begin{array}{ccl}
  t_{TCP}& =& (1-\frac{1}{3\eta})\phi, \\
  m_{TCP}& =&\sqrt{1-t_{TCP}},\\
  h_{TCP}& =&  \frac{1}{2} ln \frac{1+\sqrt{1-t_{TCP}}}{1-\sqrt{1-t_{TCP}}}+\frac{1-\eta}{1+\eta}\sqrt{1-t_{TCP}}
  \end{array}
\end{equation}
where $\phi=z_{2}J+{z_{1}J^{'}}$ and $t_{t}$ connects first and
second order transition lines. On the other hand, for $r =0.3$ and $t=t_{TCP}$
$\psi_{2}=\psi_{4}=\psi_{6}=0$ which denotes a higher order critical point (TCP) \cite{Cohen,Moreira}. Consequently, topology of the metamagnetic Ising
model phase diagram depends on the value of the ratio of the exchange interactions ($\eta=\frac{z_{1}J}{z_{2}J^{'}}$):
(i)If $r >0.3$, the transitions between the anti-ferromagnetic and
paramagnetic phases are of first order at low temperatures and
strong fields while it is of second order at higher temperatures.
The two types of transitions are connected by a tricritical point.
(ii) For  $0< r < 0.3$, the tricritical point decomposes into
a critical end point (CEP) and a bicricital end point (BCP) with a line of first order transitions in between,
separating two anti-ferromagnetic phases \cite{Cohen,Selke}, see Fig.7(a)-(b).
 The staggered magnetic susceptibility of an metamagnetic sytem is
 \begin{equation}
   \label{8}
    \chi_{s}=\lim\limits_{H_{s_{r}}\rightarrow0}\frac{\partial m_{s}}{\partial H_{s_{r}}}.
 \end{equation}
If one uses this definition and the  equations of state given in  Eq.(\ref{8}), after some algebra the staggered magnetic can  written as below
\begin{equation}
\label{9}
  \chi_{s}=\lim\limits_{H_{s_{r}}\rightarrow0}  \frac{c_{2}a_{12}-a_{22}c_{1}}{a_{21}a_{12}-a_{22}a_{11}}.
\end{equation}\\
\indent Here,
$ H_{r}=H/{J}'$ ,$H_{s_{r}}=H_{s}/{J}'$ ,
$ a_{11}, a_{12}, a_{21}, a_{22}, c_{1}$, and  $c_{2}$ are given in Appendix A. Whereas direct magnetic susceptibility ($\chi_{\; t}$) is the response function of a system  to a physical field and magnetic susceptibility can be expressed as,
\begin{equation}\label{8}
\chi_{t}=\lim\limits_{H_{s_{r}}\rightarrow0}\frac{\partial m_{t}}{\partial H_{r}}.
\end{equation}
Following the similar steps we have used in obtaining staggered magnetic susceptibility, one obtains $\chi_{\; t}$ as

\begin{equation} \label{9}
  \chi_{t}  =  \lim\limits_{H_{s_{r}}\rightarrow0} \{ \frac{b_{22}d_{1}-b_{12}d_{2}}{b_{11}b_{22}-b_{12}b_{21}}\}.
\end{equation}

\section{Results}
Fig.2(a)-(b) represents the behavior of staggered and direct magnetic susceptibilities of spin-$1/2$
metamagnetic Ising model in the neighborhood of TCP for $r=1.0$. One can see from the figure that
staggered  susceptibility ($\chi_{s}$) increases rapidly with increasing temperature and diverges at the tricritical point.
Whereas  there is a discontinuity in the direct magnetic susceptibility ($\chi_{t}$) at the TCP.
At this point we should note that  $\check{Z}$ukovic et.al. has represented a study on dilute metamagnetic Ising Model
within effective field theory \cite{Zukovic1,Zukovic2} which takes account the spin correlations.
Comparing  Fig.12 of Ref. \cite{Zukovic1}  with Fig. 1(b) of the
present Letter, one can see that our results are in accordance with the results of
effective field theory. It's important to note that, this behavior is in accordance  with the existence
of the discontinuity in  $\chi_{t}$ on the critical curve, confirming that the transition there is second-order in the
mean-field approximation \cite{lav} as well as effective field theory \cite{Zukovic1,Zukovic2}. On the other hand, Fig.3 illustrates the temperature variation of the tricritical direct magnetic susceptibility for various values of the ratio of the exchange interactions ($r$). One can see
from this figure that the amplitude of the ferromagnetic susceptibility rises considerably high values
for $r>1.78$. One of the characteristic behavior of the metamagnetic Ising model for  strong  anti-ferromagnetic
case $(r<0.3)$ is the existence of the re-entrance phenomena. One can see this fact in the phase diagram of metamagnetic Ising model for $r>0.3$.
For high values of the magnetic field the system
is in a disordered state for $T_{r}\rightarrow0$ whereas
there is a transition from disorder to order at a finite temperature.
In addition under goes another second order transition from ordered phase
to disordered phase in high temperature regime ( see Fig.7(a) and (b)).

Fig.4(a) illustrates the temperature dependence of anti-ferromagnetic susceptibility
for $H=H_{r_{DCP}}=1.994$ which corresponds to the double critical end point(DCP)
of the spin-$1/2$ Ising model for $r=0.2$. Here $T_{r_{N_{1}}}$ at which staggered
susceptibility diverges denotes
the Neel temperature at which system undergoes a second order transition from
paramagnetic phase to antiferromagnetic phase. Whereas the second divergence of the
staggered susceptibility takes place at  $T_{r_{DCP}}$. In addition one can clearly
see that there exist an non-critical maximum  at the ordered phase . This maximum
corresponds to a anomaly in the multicritical behavior of
iron group dihalides.

It is important to emphasize that  Selke has
reported that
there is two lines of anomalies  in  the field-temperature phase diagram
of the spin-$1/2$ Ising model with in mean field theory
at which which the temperature derivative of the total magnetization
exhibits, at fixed field value a maximum below the transition
point, as exemplified in Fig. 3 of Ref.\cite{Selke}.
In this study the anomalies are related to the competing ordering tendencies of the external  field and the interlayer couplings in a metamagnetic crystal. We should also note that there are experimental data which emphasizes the anomalies  for quite some time \cite{Cohen}.
In addition, there have been  various  experimental study on the field-induced Griffiths phase in Ising type metamagnets such as $FeBr_{2}$, $FeCl_{2}$ and $Fe_{1-x}Zn_{x}F_{2}$ \cite{Binek1,Binek2}. Fig.4(b) exhibits the temperature variation of ferromagnetic susceptibility for  $H=H_{r_{DCP}}$.
 In this case the signature of the second order transition from
paramagnetic phase to antiferromagnetic phase is a discontinuity in the direct magnetic susceptibility which is in accordance with
the literature \cite{lav}. In addition, there exist a special multicritical point which separates
the two different anti- ferromagnetic phases (AFI and  AFII). This special
continuous phase transition is of fourth order and direct magnetic susceptibility
represents a discontinuity at the double critical end point.
In addition direct susceptibility also represent a
discontinuity at $T_{r_{N_{2}}}$ which corresponds to a regular
critical point from antiferromagnetic phase to paramagnetic phase.
Fig.6(a) shows the temperature dependencies of staggered and
direct susceptibilities of the  spin-$1/2$ metamagnetic system for $r=0.2$
and $H=H_{r_{CEP}}=1.99176$. At this value of the reduced physical magnetic field
the system under goes two phase transitions of different character.
The first transition is the CEP which is also of fourth order \cite{Cohen} and takes place
between disordered phase at lower temperatures and anti ferromagnetic phase at higher temperature regime.
One can easily observe from  Fig.6 that the staggered susceptibility diverges at CEP whereas  the direct magnetic
susceptibility shows a discontinuity. Similar to the anomaly at $H=H_{r_{DCP}}$ , both $\chi_{\; t}$ and $\chi_{\; s}$
makes non-critical maximums in the antiferromagnetic phase. In Fig.6 (a) and (b) we have given the temperature variances of
the magnetic response functions of the system for different constant reduced physical field values. One can see from these figures that
the broad maximum in the ordered phase declines with decreasing the amplitude of the physical external magnetic field. Finally, the line of anomalies in the staggered and direct susceptibilities is depicted in Fig.7. Here $(T-H)_{\chi}$ denotes the field and  temperature values which both the staggered and direct susceptibilities exhibit a broad maximum in the ordered phase as exemplified in Figs.4 and 5. Unlike the anomalies discussed by Selke the broad maximum does not diverges as one approaches the double critical endpoint. Further, the anomalies the magnetic response functions of the metamagnetic Ising system disappears for the case  $r\geq 0.3$ where the critical end point and the double critical end points emerges to a tricritical point.  In this case there is no re-entrance in the phase diagram.

\section{Conclusions and Discussions}
In this paper the magnetic response of iron group  dihalides in a field with weak ferromagnetic
intralayer interactions and highly coordinated antiferromagnetic interlayer couplings are studied  within mean field approximation.
The expressions that describe the staggered (anti- ferromagnetic) and direct (ferromagnetic)
susceptibilities are derived by making use of mean field theory. The findings of this study can be
summarized as follows: the direct susceptibility exhibits
discontinuity not only at the second order transition point but also
at multicritical points such as TCP, CEP, and DCP. In addition,
the both magnetic response functions of the metamagnetic Ising model
exhibits non-critical maximums in the ordered phase at the  region
of the $H_{r}-T_{r}$ in where the system shows re-entrance phenomena.

\section{Acknowledgements}
This work was supported by the Scientific and Technological Research Council of Turkey (TUBITAK), Grant No. 109T721.
In addition authors thank A.N. Berker for valuable discussions, Sabanci University and Massachusetts Institute of Technology.
\newpage
\section{Appendix A}
The coefficients $ a_{11}, a_{12}, a_{21}, a_{22}, c_{1}$, and  $c_{2}$  in Eq.(\ref{9}) are defined as follows:
\begin{equation}
   \label{61}
    \begin{array}{lcl}
  a_{11}&=& 1-\left(1-tanh\left(\frac{-2\left(m_{t}-m_{s}\right)+4r(m_{t}+m_{s})+H_{r}+H_{s_{r}}}{T_{r}}\right)^{2}\right)(2+4r)T_{r}^{-1},\\
  \\
  a_{12}&=& 1-\left(1-tanh\left(\frac{-2(m_{t}-m_{s})+4r(m_{t}+m_{s})+H_{r}+H_{s_{r}}}{T_{r}}\right)^{2}\right)(-2+4r)T_{r}^{-1},\\
  \\
  a_{21}&=& -1+\left(1-tanh\left(\frac{-2(m_{t}+m_{s}+4r(m_{t}-m_{s})+H_{r}-H_{s_{r}}}{T_{r}}\right)^{2}\right)(2+4r)T_{r}^{-1},\\
  \\
  a_{22}&=& 1-\left(1-tanh\left(\frac{-2(m_{t}+m_{s})+4r(m_{t}-m_{s})+H_{r}-H_{s_{r}}}{T_{r}}\right)^{2}\right)(-2+4r)T_{r}^{-1},\\
  \\
  c_{1}&=& \left(1-tanh\left(\frac{-2(m_{t}-m_{s})+4r(m_{t}+m_{s})+H_{r}+H_{s_{r}}}{T_{r}}\right)^{2}\right)T_{r}^{-1},\\
  \\
  c_{2}&=& \left(-1+tanh\left(\frac{-2(m_{t}+m_{s})+4r(m_{t}-m_{s})+H_{r}-H_{s_{r}}}{T_{r}}\right)^{2}\right)T_{r}^{-1},\\
  \end{array}
\end{equation}
\newpage
\section{Appendix B}
The coefficients $ b_{11}, b_{12}, b_{21}, b_{22}, d_{1}$, and  $d_{2}$  in Eq.(\ref{9}) are defined as follows:
\begin{equation}
\label{66}
    \begin{array}{lcl}
  b_{11}&=&1-\left(1-tanh\left(\frac{-2(m_{t}-m_{s})+4r(m_{t}+m_{s})+H_{r}+H_{s_{r}}}{T_{r}}\right)^{2}\right)(2+4r)T_{r}^{-1},\\
  \\
  b_{12}&=&1-\left(1-tanh\left(\frac{-2(m_{t}-m_{s})+4r(m_{t}+m_{s})+H_{r}+H_{s_{r}}}{T_{r}}\right)^{2}\right)(-2+4r)T_{r}^{-1},\\
  \\
  b_{21}&=&-1+\left(1-tanh\left(\frac{-2(m_{t}+m_{s})+4r(m_{t}-m_{s})+H_{r}-H_{s_{r}}}{T_{r}}\right)^{2}\right)(2+4r)T_{r}^{-1},\\
  \\
  b_{22}&=&1-\left(1-tanh\left(\frac{-2(m_{t}+m_{s})+4r(m_{t}-m_{s})+H_{r}-H_{s_{r}}}{T_{r}}\right)^{2}\right)(-2+4r)T_{r}^{-1},\\
  \\
  d_{1}&=&\left(1-tanh\left(\frac{(-2(m_{t}-m_{s})+4r(m_{t}+m_{s})+H_{r}+H_{s_{r}})}{T_{r}}\right)^{2}\right)T_{r}^{-1},\\
  \\
  d_{2}&=&\left(1-tanh\left(\frac{(-2(m_{t}+m_{s})+4r(m_{t}-m_{s})+H_{r}-H_{s_{r}})}{T_{r}}\right)^{2}\right)T_{r}^{-1}.
    \end{array}
\end{equation}

\newpage
\clearpage

\begin{figure}[tbp]
\begin{center}
 \includegraphics[width=6cm,height=8cm,angle=0,bb=0 0 1000 1000]{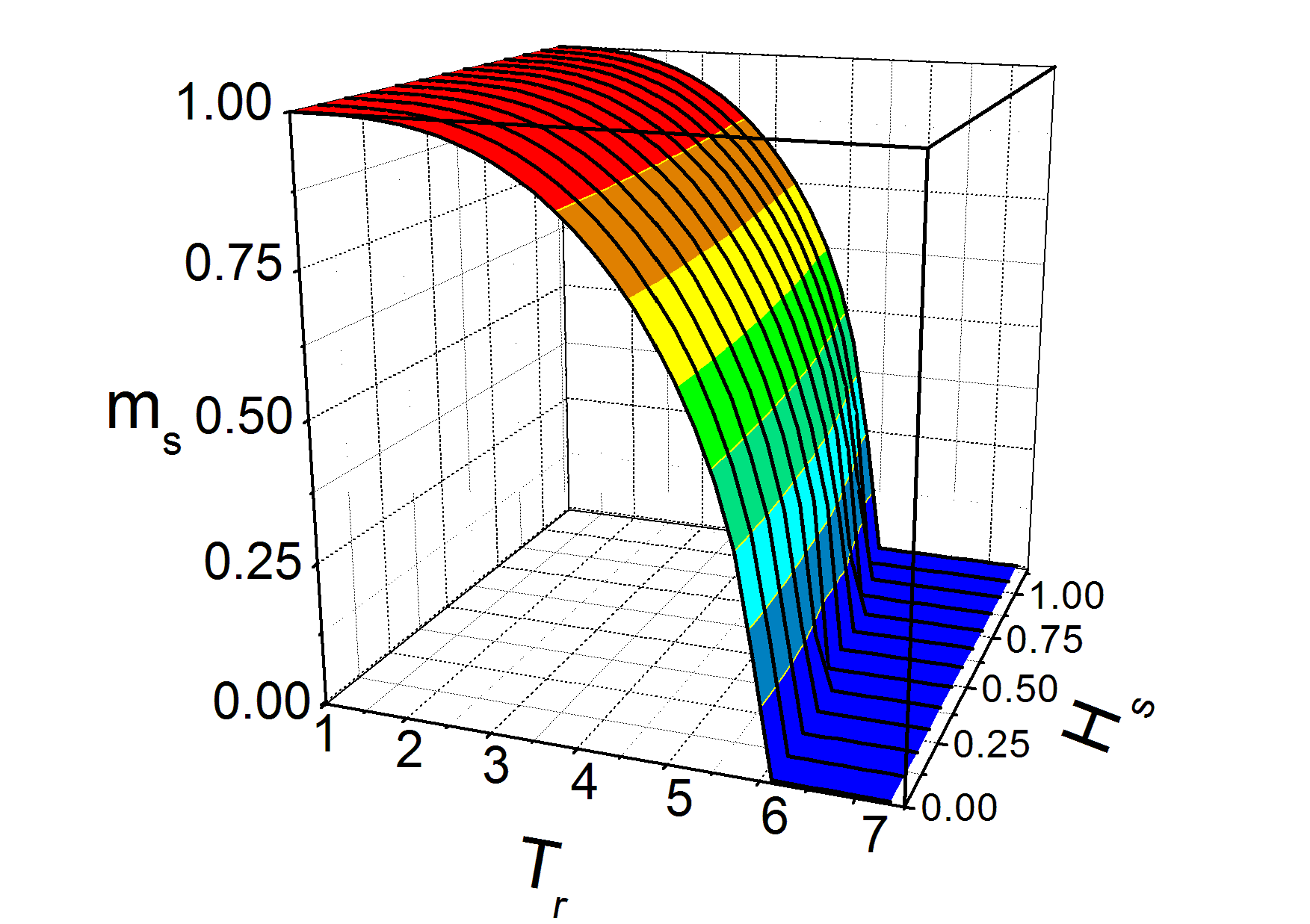}
  \includegraphics[width=6cm,height=8cm,angle=0,bb=0 0 1000 1000]{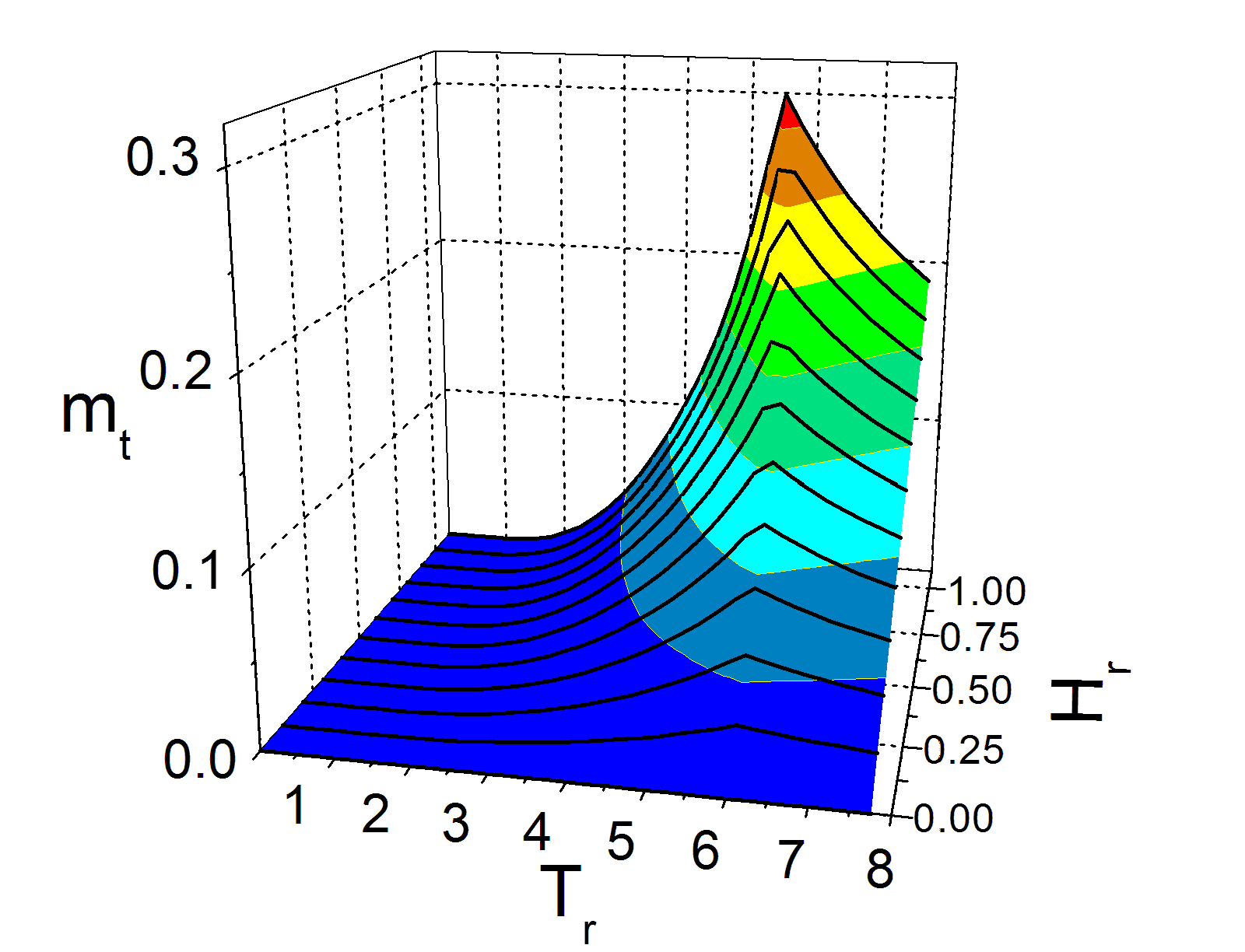}
 \caption{Variation of the antiferromagnetic (staggered) magnetization in the reduced
field- reduced temperature plane for a second order phase transition.}
\end{center}
\end{figure}

\begin{figure}[tbp]
\begin{center}
 \includegraphics[width=6cm,height=8cm,angle=0,bb=0 0 1000 1000]{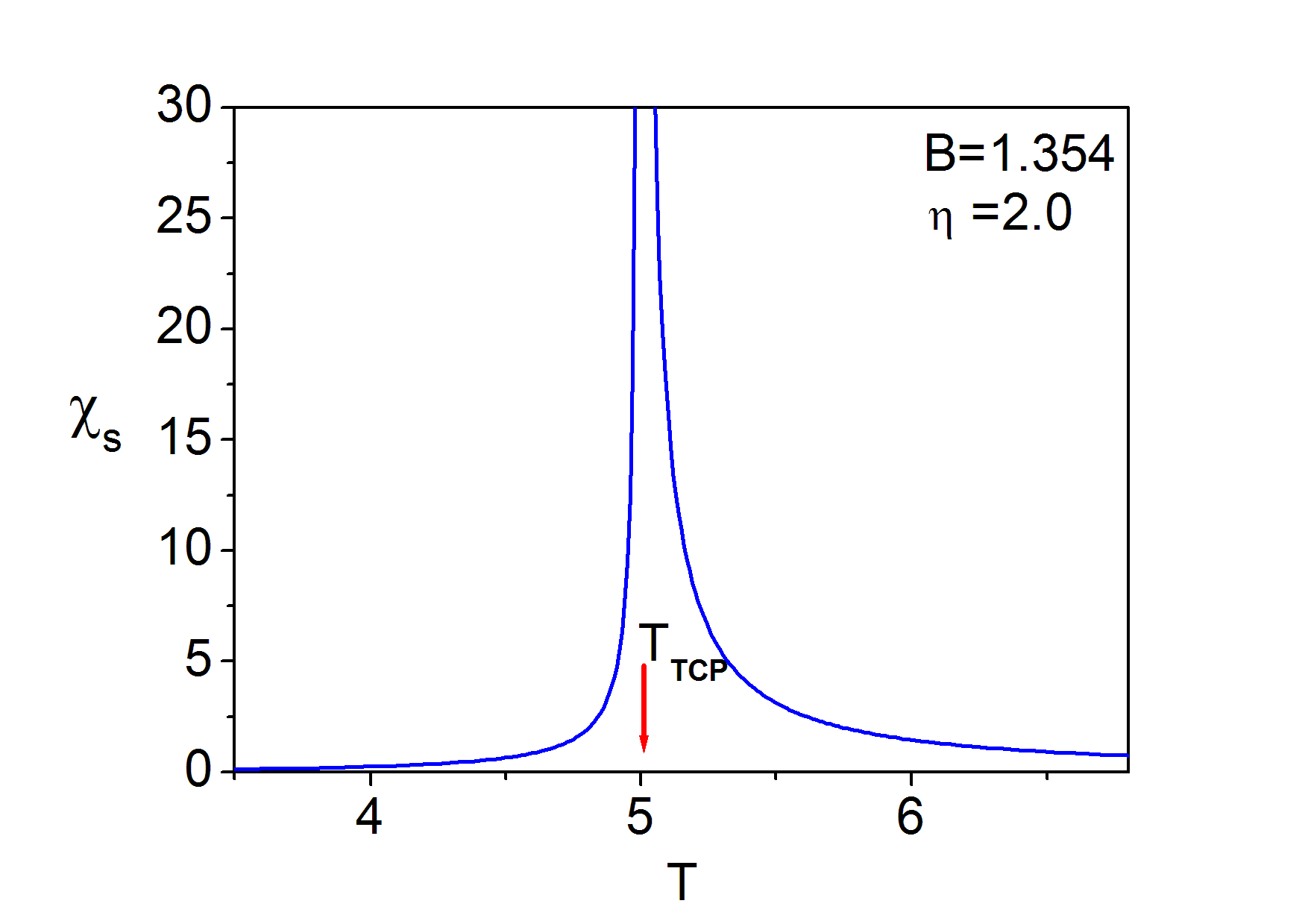}
  \includegraphics[width=6cm,height=8cm,angle=0,bb=0 0 1000 1000]{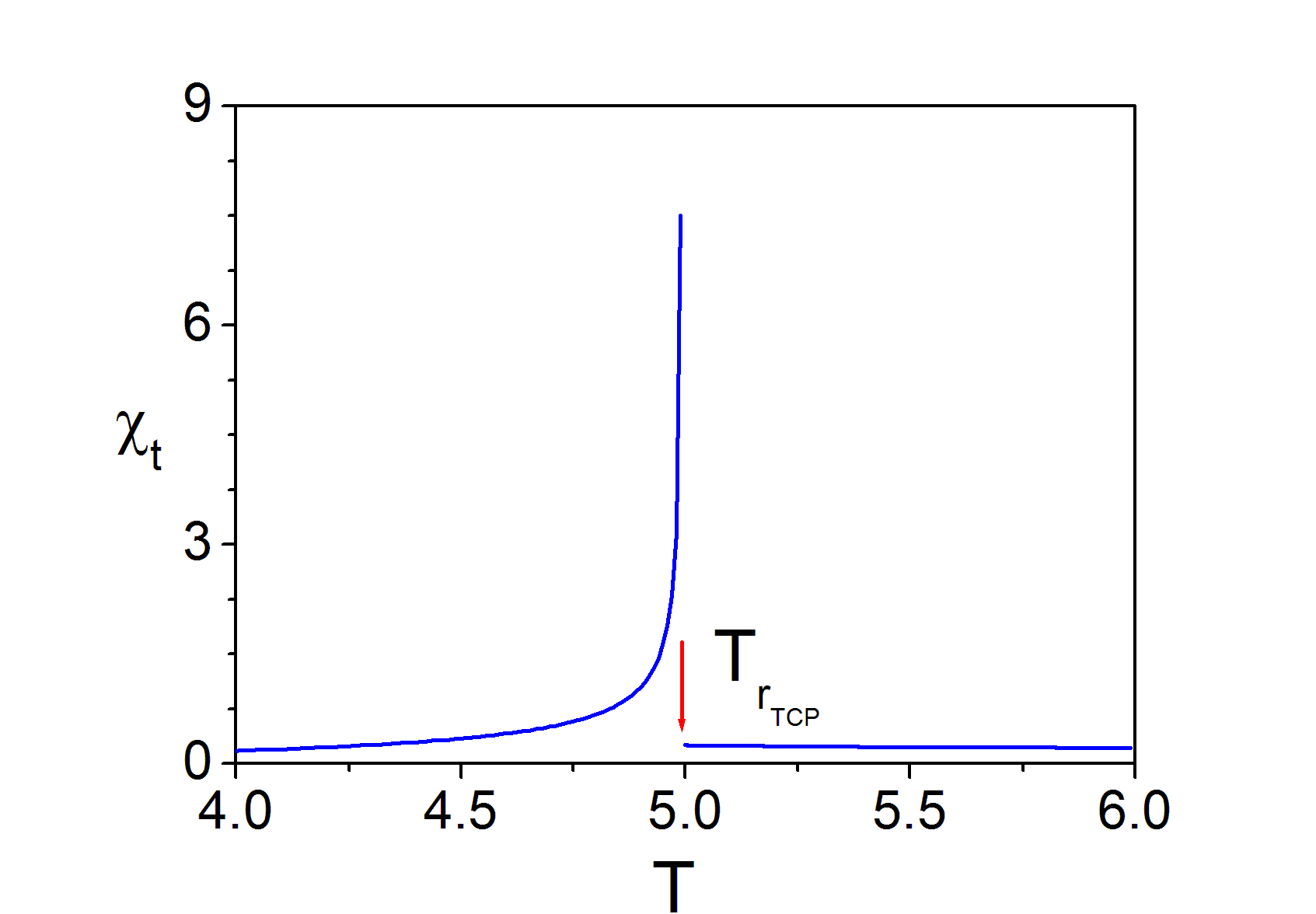}
 \caption{Behaviors of staggered (a) and magnetic (b) susceptibilities each as a function of reduced temperature for $H_{r}=\frac{H}{J^{'}}=1.354$ and $r=J/J^{'}=1.0.$ For these values,  the system has a tricritical point (TCP). And also, staggered susceptibility increases rapidly with increasing temperature and diverges as the temperature approaches to the TCP.}
\end{center}
\end{figure}

\begin{figure}[tbp]
\begin{center}
\includegraphics[width=6cm,height=8cm,angle=0,bb=0 0 1000 1000]{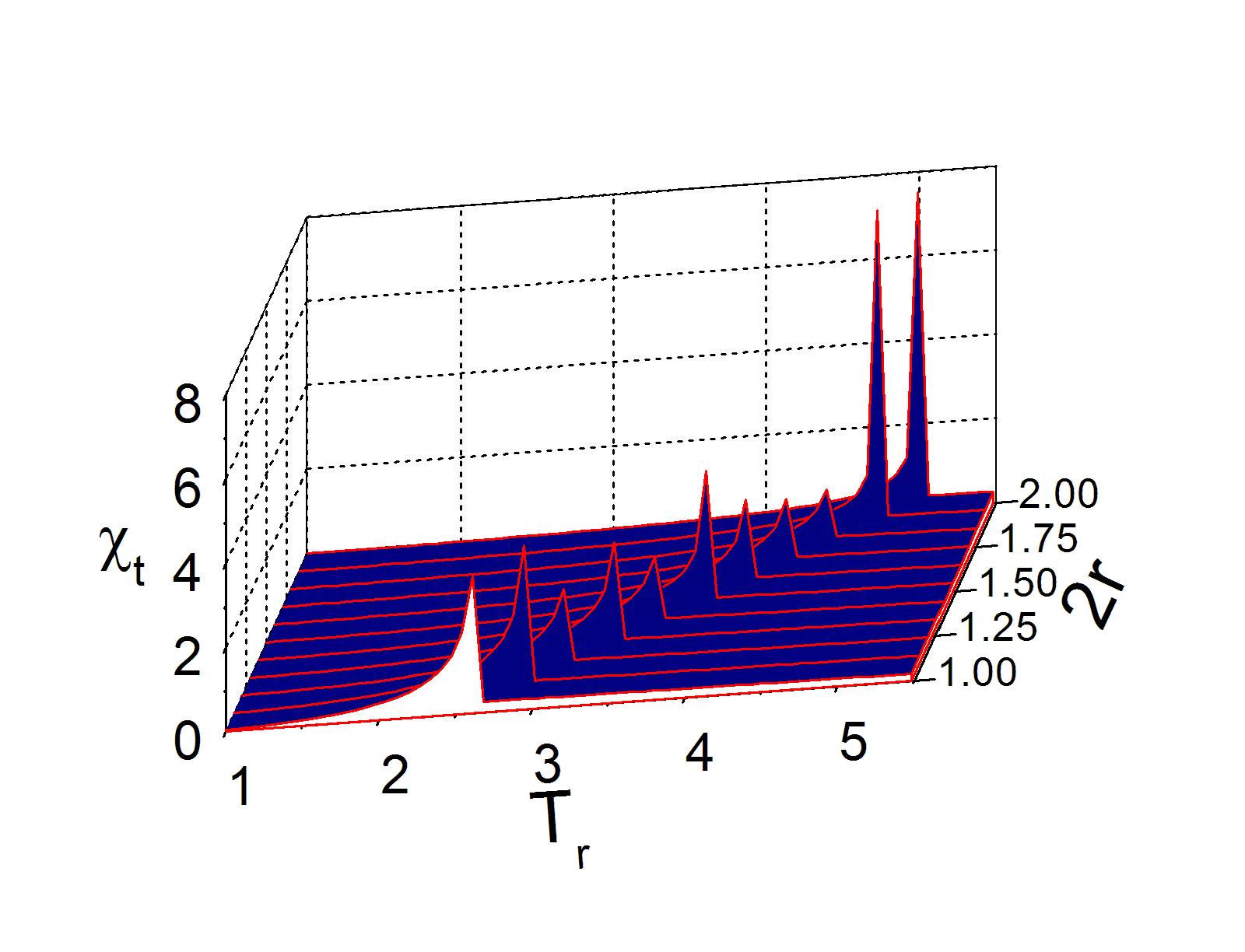}
 \caption{The  temperature variation of the tricritical
direct magnetic susceptibility for various values of the ratio of the exchange interactions. Here the arrows  illustrate the phase transition temperatures.}
\end{center}
\end{figure}

\begin{figure}[tbp]
\begin{center}
 \includegraphics[width=6cm,height=8cm,angle=0,bb=0 0 1000 1000]{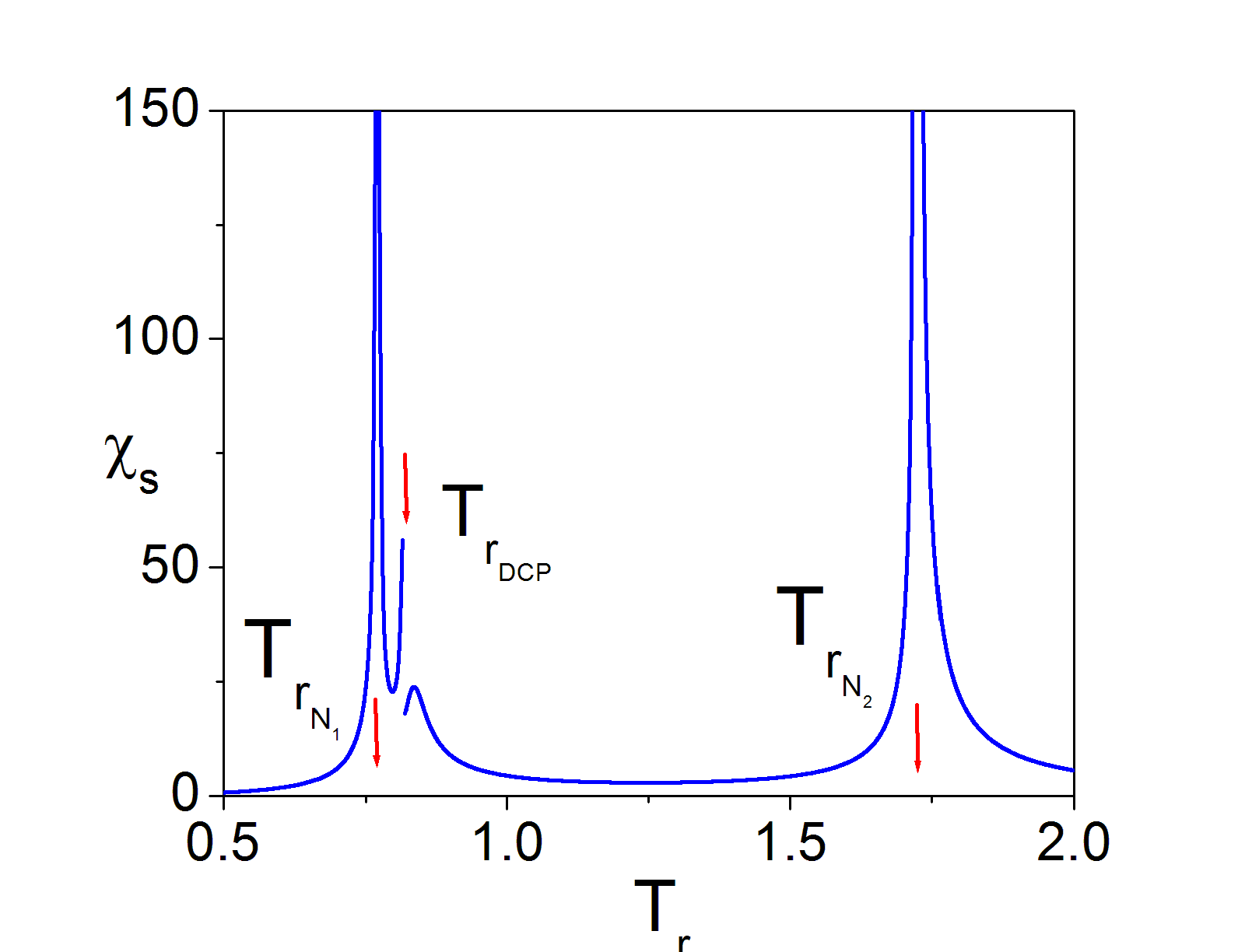}
  \includegraphics[width=6cm,height=8cm,angle=0,bb=0 0 1000 1000]{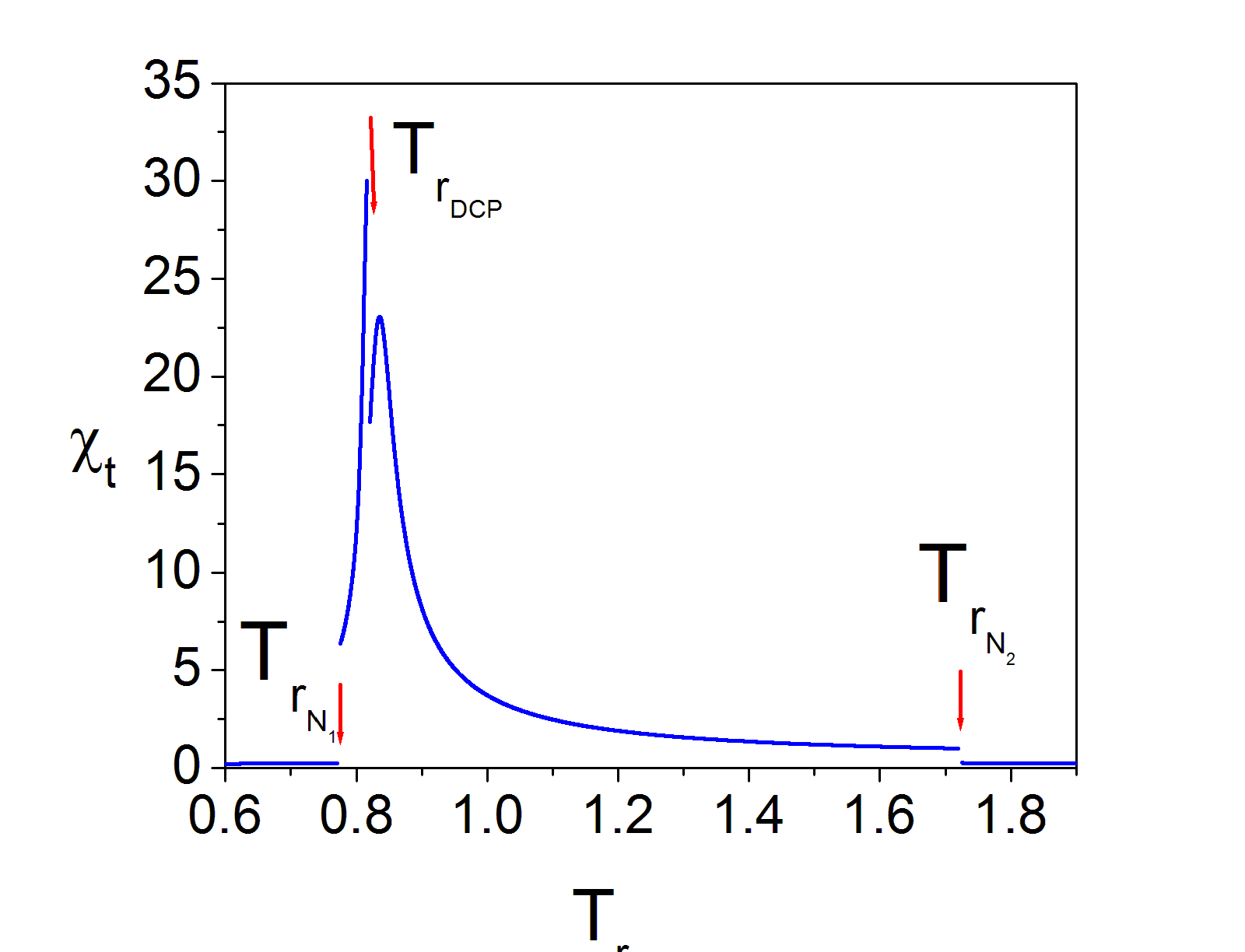}
 \caption{The temperature dependencies of staggered  and total susceptibilities in the neighborhood of  double critical  point (DCP) and second order phase transition point which takes place for the value of the reduced magnetic field $H_{r_{DCP}}=1.994$ for $r=0.2$. Here the arrows  illustrate the phase transition temperatures.}
\end{center}
\end{figure}

\begin{figure}[tbp]
\begin{center}
 \includegraphics[width=6cm,height=8cm,angle=0,bb=0 0 1000 1000]{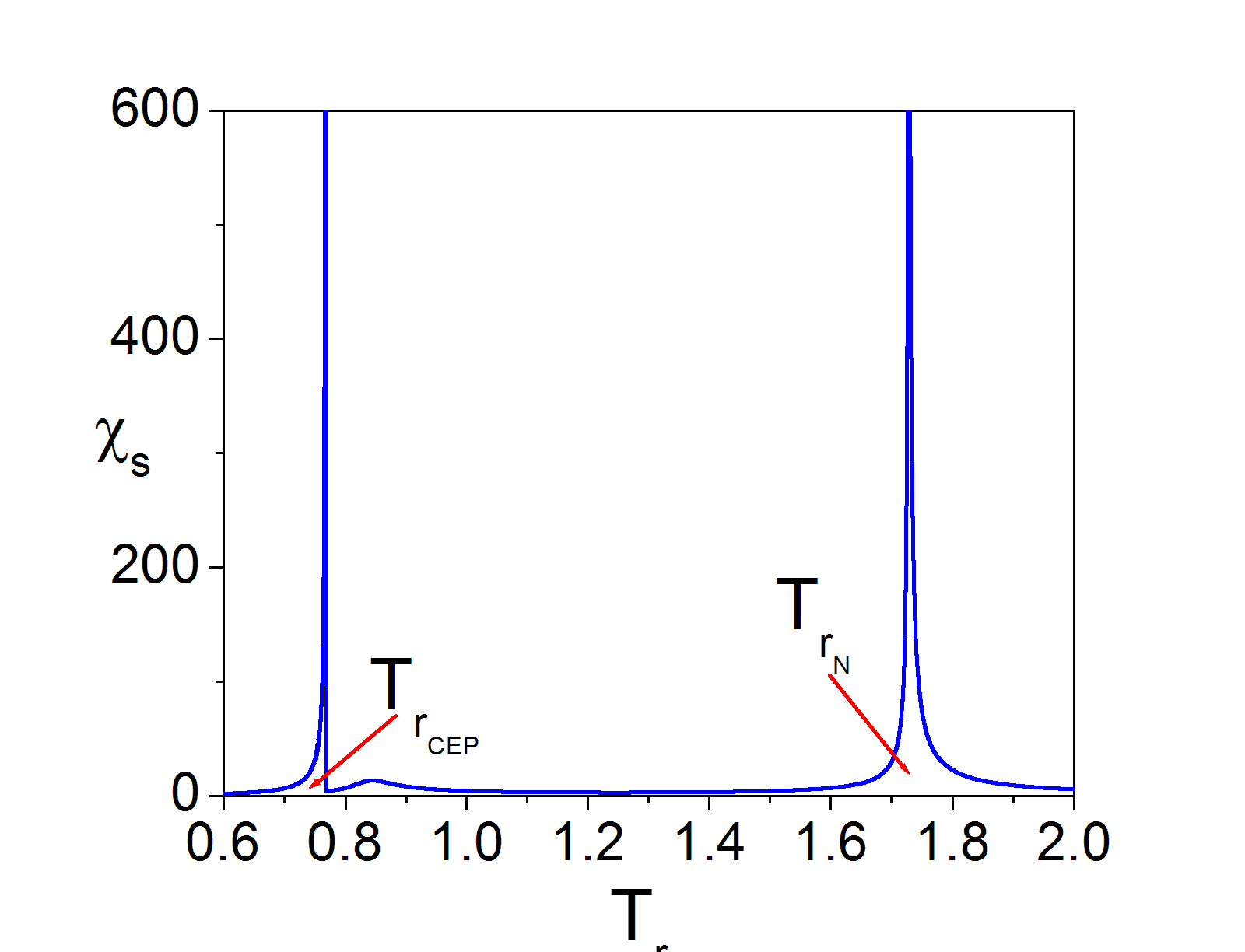}
  \includegraphics[width=6cm,height=8cm,angle=0,bb=0 0 1000 1000]{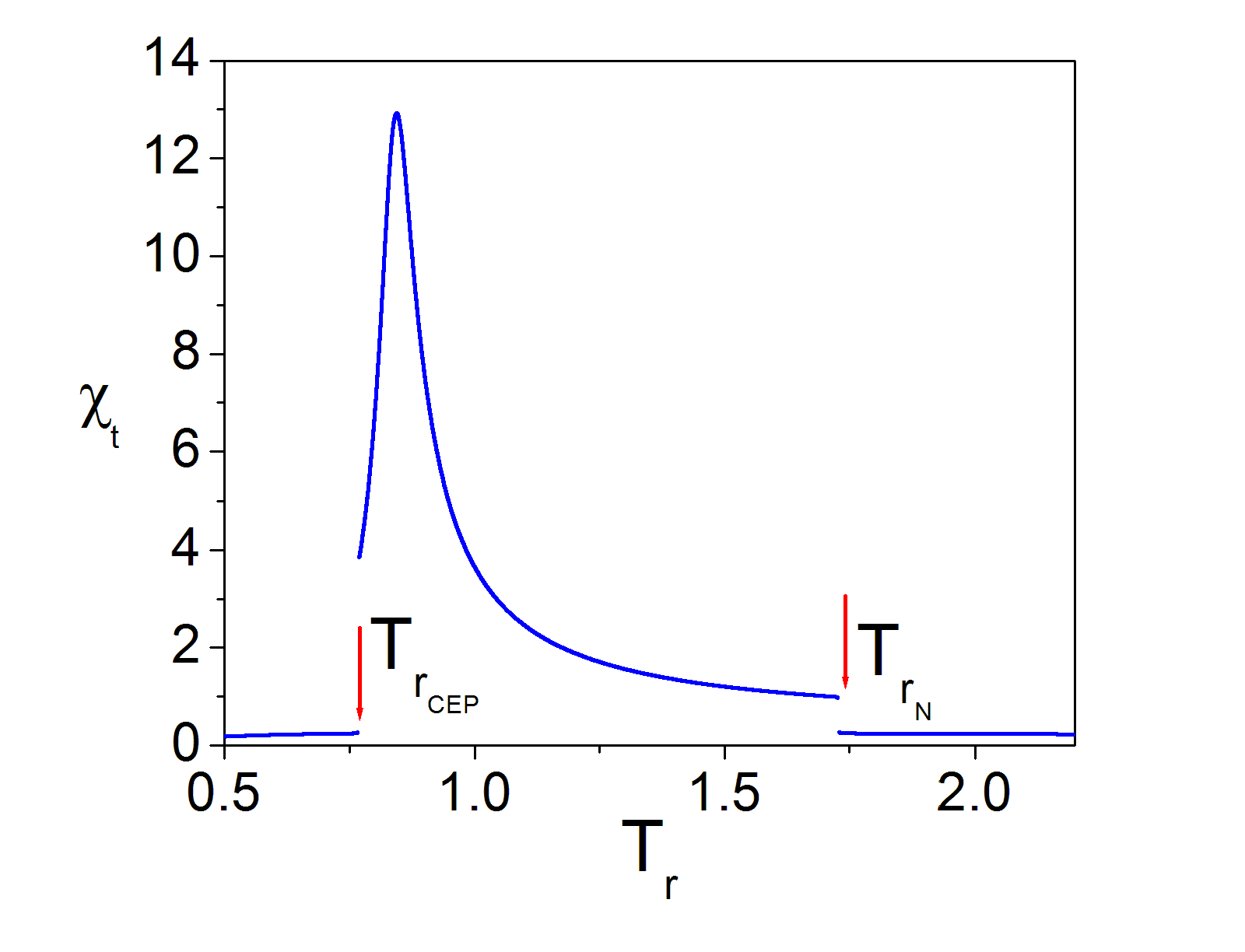}
 \caption{The temperature dependencies of staggered  and total susceptibilities in the neighborhood of  the critical end point (CEP) and second order phase transition point which takes place for the value of the reduced magnetic field $H_{r_{CEP}}=1.99176$ for $r=0.2$. Here the arrows  illustrate the phase transition temperatures.}
\end{center}
\end{figure}

\begin{figure}[tbp]
\begin{center}
 \includegraphics[width=6cm,height=8cm,angle=0,bb=0 0 1000 1000]{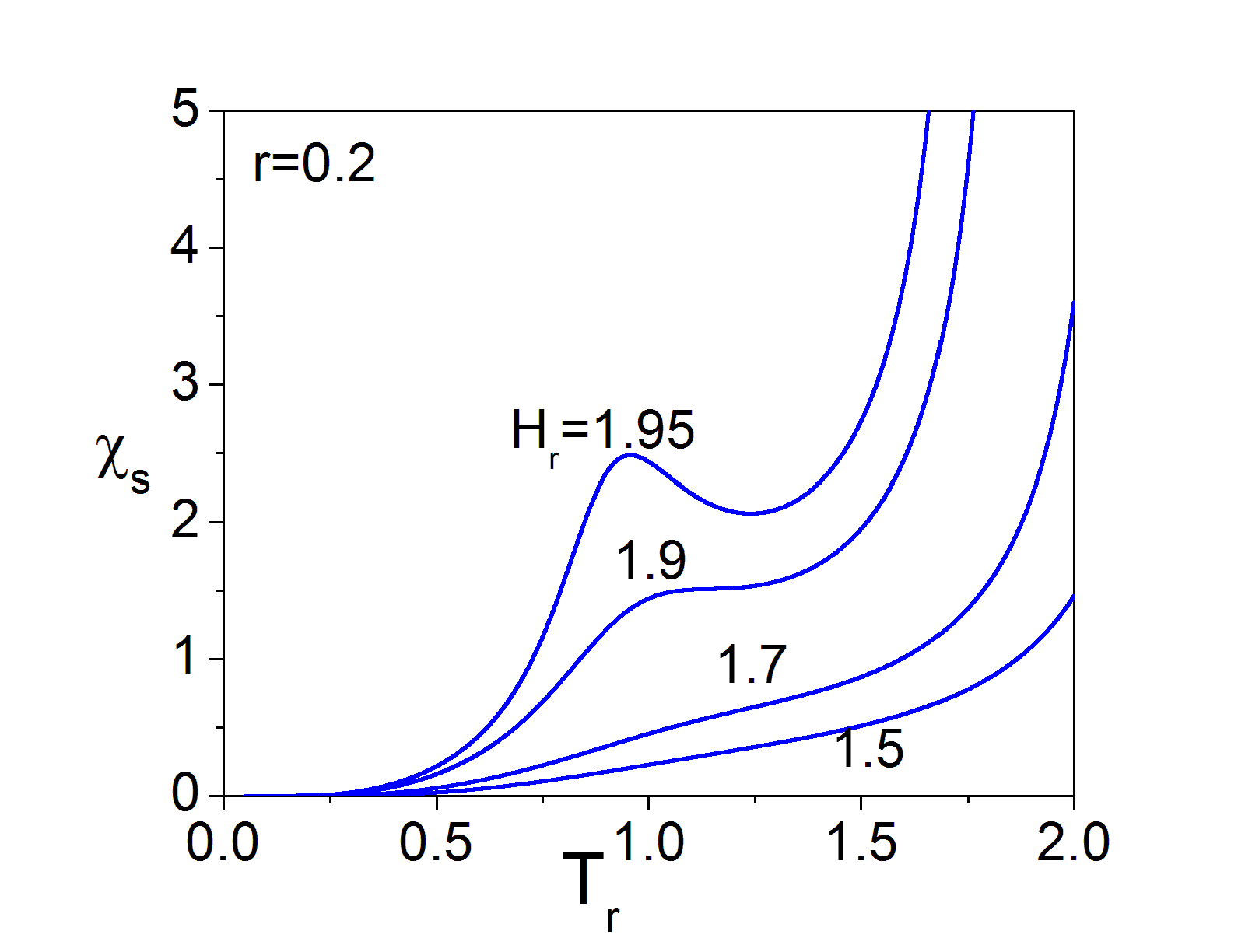}
  \includegraphics[width=6cm,height=8cm,angle=0,bb=0 0 1000 1000]{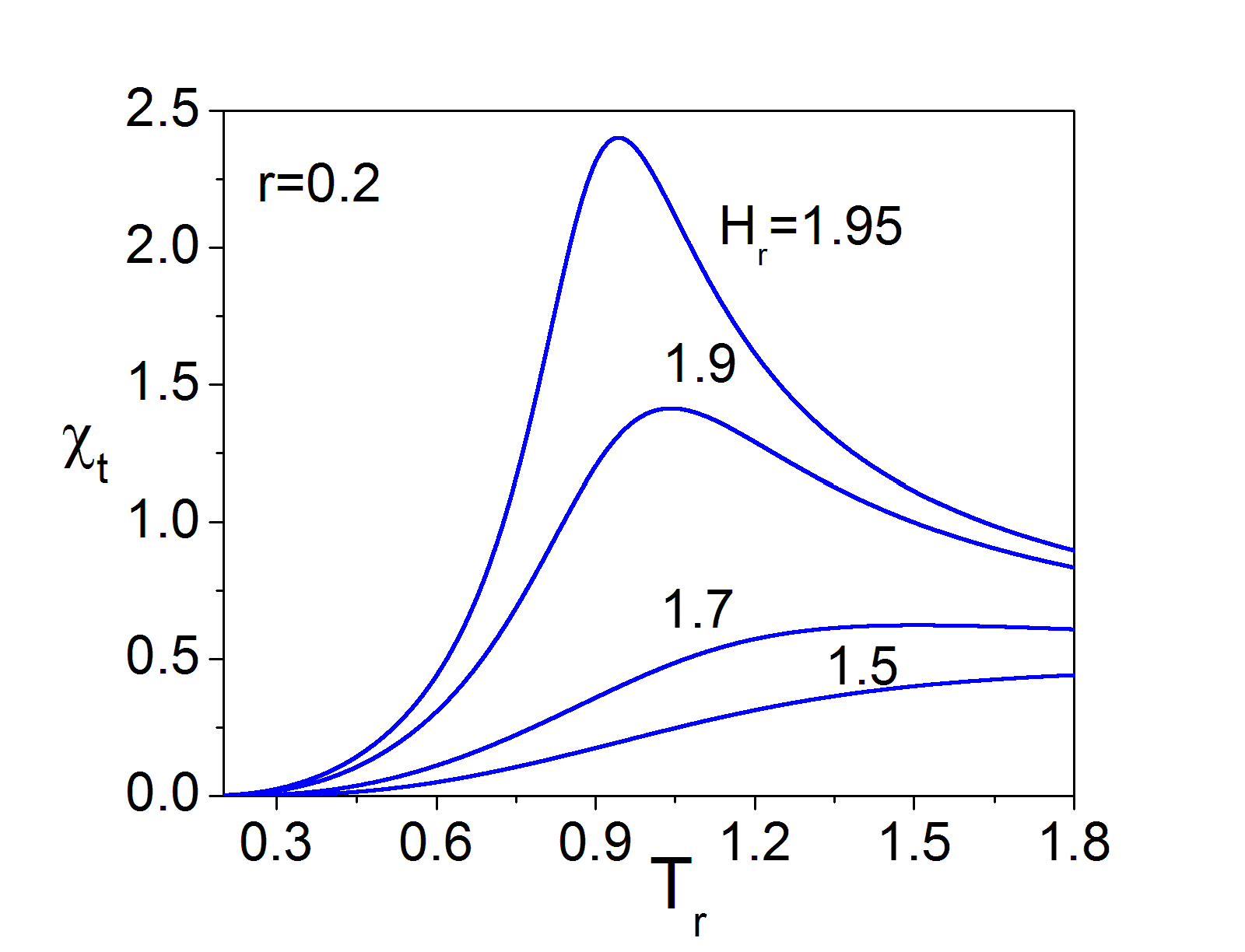}
 \caption{The behavior of (a) the staggered susceptibility $\chi _{s}$ (b) the direct magnetic susceptibility $\chi _{t}$ as a function
of the reduced temperature, where $T_{r}=\frac{k_{B}T}{J^{'}}$ is
 for several values of reduced field, $H_{r}=\frac{H}{J^{'}}$.}
\end{center}
\end{figure}

\begin{figure}[tbp]
\begin{center}
 \includegraphics[width=6cm,height=8cm,angle=0,bb=0 0 1000 1000]{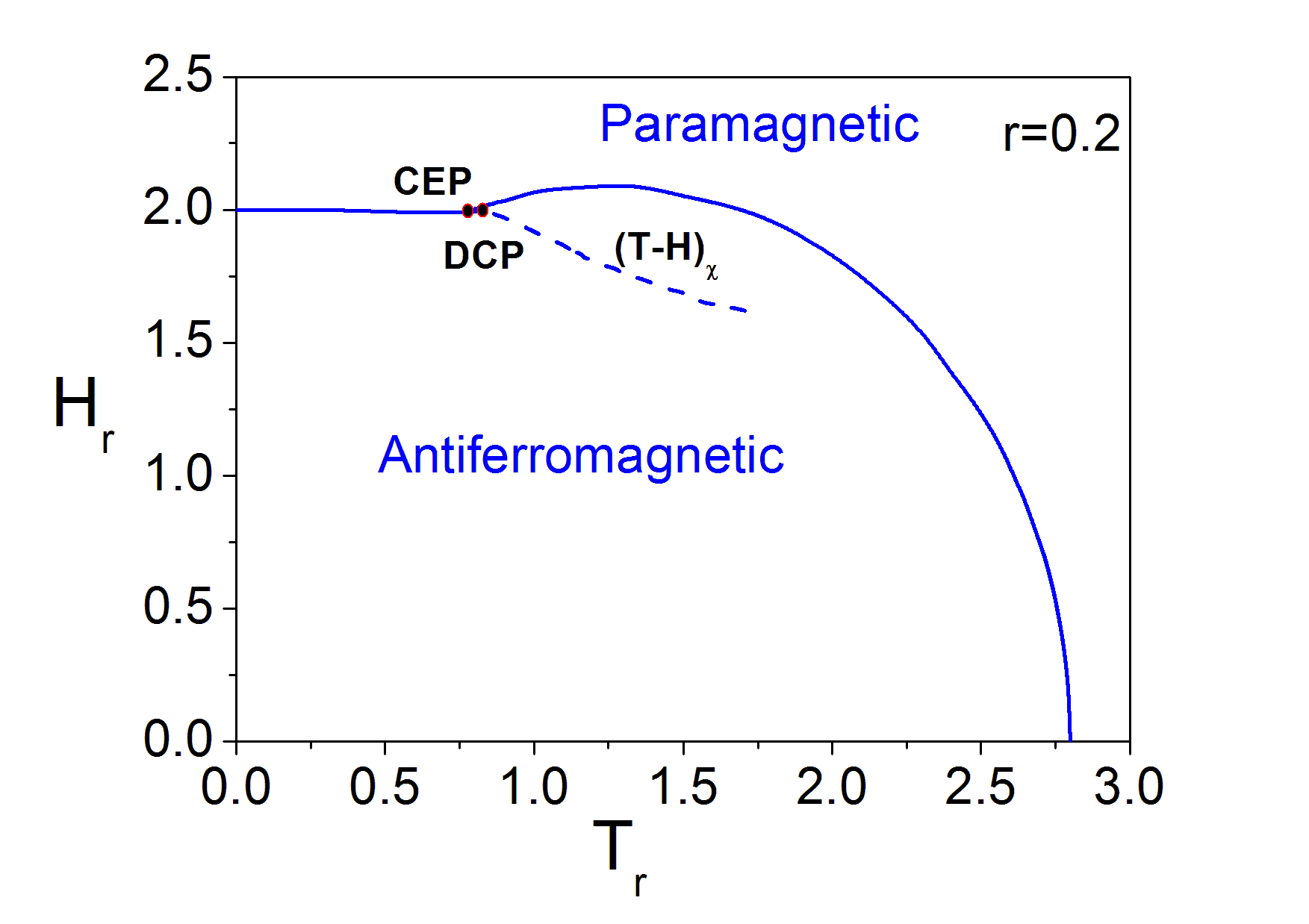}
  \includegraphics[width=6cm,height=8cm,angle=0,bb=0 0 1000 1000]{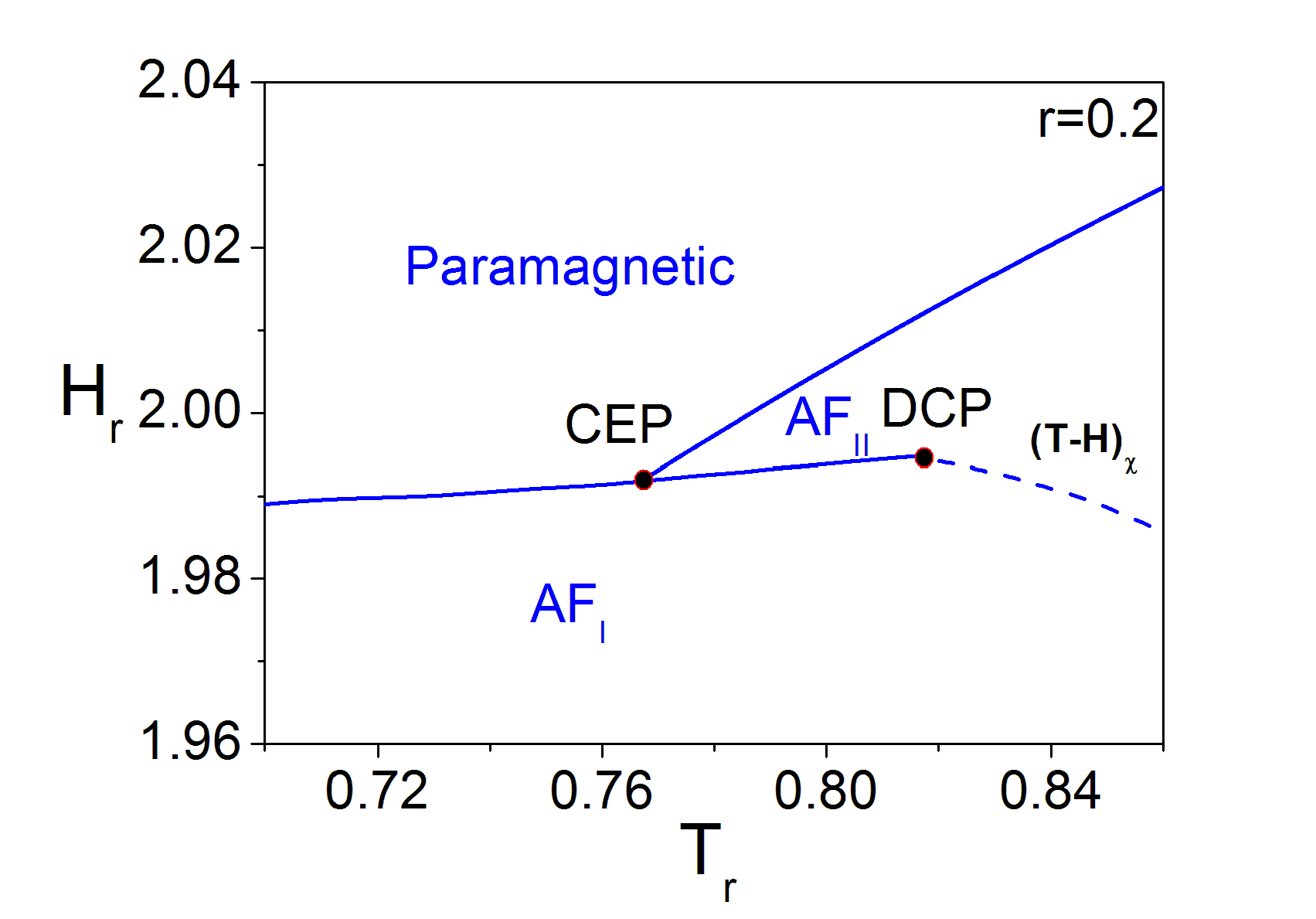}
 \caption{The calculated mean field phase diagram of the metamagnetic Ising model for $r=\frac{J}{J^{'}}=0.2$ in the temperature- field plane.
(b) Detailed phase diagram in the neighborhood of the critical end point (CEP) and the double critical end point (DCP).
The dashed lines denote the anomalies in the staggered and direct susceptibilities.}
\end{center}
\end{figure}

\begin{thebibliography}{99}
\bibitem{Taufour}   V. Taufour, D. Aoki, G. Knebel, and J. Flouquet, Phys. Rev. Lett. 105 (2010) 217201.
\bibitem{Hankey}    A. Hankey, T. S. Chang, and H. E. Stanley, Phys. Rev.A 9 (1974) 2573.
\bibitem{Ginzberg}  B. Ginzberg; S. Bergerman; D. H. Kurlat, Phys. and Chem. of Liq. 27 (1994) 83.
\bibitem{Garland}   C. W. Garland and B. B. Weiner, Phys. Rev.B  3 (1973) 1634.
\bibitem{Riedel}    E. K. Riedel, Phys. Rev. Lett.   28 11 (1972).
\bibitem{Peercy}    P. S. Peercy, Phys. Rev. Lett. 35 (1975) 1581 ; V. H. Schmidt, Bull. Am. Phys. Soc. 19 (1974) 649.
\bibitem{Leidl}     R. Leidl, H. W. Diehl, Phys. Rev. B 57 (1998) 1908.
\bibitem{Iwata}     M. Iwata, Z. Kutnjak, Y. Ishibashi, and R. Blinc, J. Phys. Soc. Jpn. 77 (2008) 065003.
\bibitem{Konynen}   P. H. van Konynenburg and R.L. Scott, Philos. Trans. R. Soc. Ser. A 298 (1980) 495.
\bibitem{Demko}     L. Demko, I. Kezsmarki, G. Mihaly, N. Takeshita, Y. Tomioka, and Y. Tokura, Phys. Rev. Lett. 101 (2008) 037206.
\bibitem{Kauf}      M. Kaufman, P.E. Klunzinger,  A. Khurana, Phys. RevB. 34 (1986) 4766 .
\bibitem{Selke}     W. Selke, Z. Phys. B 101 (1996) 145.
\bibitem{Wild}      N. B. Wilding, Phys. Rev. Lett. 78 (1997) 1488.
\bibitem{Fisher1ve2}M. E. Fisher and P.J. Upton, Phys. Rev. Lett. 65 (1990) 2402; 65 (1990) 3405.
\bibitem{Fisher3}   M. E. Fisher and M.C. Barbosa, Phys. Rev. B 43 (1991) 11177; 43 (1991) 10635.
\bibitem{Barbosa}   M. C. Barbosa, Phys. Rev. B 45 (1992) 5199.
\bibitem{Poot}      W. Poot and T.W. de Loos, Phys. Chem. Chem. Phys. 1 (1999) 4923.
\bibitem{Wolf}      W. P. Wolf, Braz. J. Phys. 30 (2000) 794.
\bibitem{Katsumata} K. Katsumata, H. Aruga Katori, S.M. Shapiro, and G. Shirane, Phys. Rev. B 55 (1997) 11466.
\bibitem{Cohen}     J. M. Kincaid and E. G. D. Cohen, Phys. Reports C 22 (1975) 57.
\bibitem{Stryj}     E. Stryjewski and N. Giordano, Adv. Phys. 26 (1977) 487.
\bibitem{Wang}      Y.-L. Wang and J.D. Kimel, J. Appl. Phys. 69 (1991) 6176.
\bibitem{Kimel}     J. D. Kimel, S. Black, P. Carter, and Y.-L. Wang, Phys. Rev. B 35 (1987) 3347.
\bibitem {Plascak}  J. A. Plascak and D. P. Landau, Phys. Rev. E 67 (2003) 015103R .
\bibitem{Alessandrini}V. A. Alessandrini, H.J. de Vega, and F. Schaposnik, Phys. Rev.B. 12 (1975) 5034 .
\bibitem{Landau} D. P. Landau, Phys. Rev.B 14 (1976) 4054 .
\bibitem{Nelson}  D. P. Landau and M.E. Fisher, Phys. Rev.B 11, 1030 (1975); ibid. 12 (1975) 263.
\bibitem{Barkowiak} M. Barkowiak, J.A. Henderson, J. Oitmaa, P.E. de Brito, Phys. Rev.B 51 (1995) 14077 .
\bibitem{Li} W. Li, S.-S.Gong, Y. Zhao, S.-J. Ran, S. Gao, and G. Su, Phys. Rev. B 82 (2010) 134434 .
\bibitem {He} Z. He and Y. Ueda, Phys. Rev. B 77 (2008) 052402.
\bibitem {Millis} A. J. Millis, C. Lampropoulos, S. Mukherjee, and G. Christou, Phys. Rev. B, 82 (2010) 174405 .

\bibitem{Giordano} E. Stryjewski, M. Giordano,  Adv. Phys. 26 (1977) 487.
\bibitem{Zhou} H.D. Zhou, E. S. Choi, Y.J. Jo, L. Balicas, J. Lu, L. L. Lumata, R. R. Urbano, P. L. Kuhns, A. P. Reyes, J. S. Brooks, R. Stillwell, S. W. Tozer, C.R. Wiebe, J. Whalen, and T. Siegrist, Phys. Rev.B 82 (2010) 054435 .
\bibitem{kalita} V. M. Kalita, A. F. Lozenko, S. M. Ryabchenko, and P. A. Trotsenko, Low Temp. Phys. 31 (2005) 794.
\bibitem{fujita} T. Fujita, A. Ito and K. Ôno, J. Phys. Soc. Jpn, 27 (1969) 1143.

\bibitem{Landau72} D. P. Landau, Phys. Rev. Lett. 28 (1972) 449.
\bibitem{Arora} B. L. Arora, D. P. Landau, AIP Conf. Proc. 10 (1973) 870.
\bibitem{Harbus1} F. Harbus, H. E. Stanley, Phys. Rev. B 8 (1973) 1156.
\bibitem{Harbus2} F. Harbus, H. E. Stanley, Phys. Rev. B 8 (1973) 1141.
\bibitem{Selke} W. Selke, Z. Phys. B 101, 145 (1996).
\bibitem{lav} D. A. Lavis, G. M. Bell, Statistical Mechanics of Lattice
Systems 1, Berlin, Springer, (1999).
\bibitem{Gul} G. Gulpinar, D. Demirhan, F. Buyukkilic, Phys. Rev. E 75, (2007) 021104.
\bibitem {Moreira} A. F. S. Moreira, W. Figueiredo, and V. B. Henriques, Eur. Phys. J. B. 27, (2002) 153.
\bibitem{Hernandez} L. Hernandez, H. T. Diep, D. Bertrand, Phys. Rev. B 47
(1993) 2602.
\bibitem{Onyszkiewicz1} Z. Onyszkiewicz, A. Wierzbicki, Physica B 151 (1988) 462.
\bibitem{Onyszkiewicz2} Z. Onyszkiewicz, A. Wierzbicki, Physica B 151 (1988) 475.

\bibitem{James} M. James, J. Phys. Chem. Solids 61 (2000) 1865.
\bibitem{Kleemann} W. Kleemann, H. A. Katori, T. Kato, Ch. Binek, K. Katsumata, Europhys. Lett. 55 (2001)
732.
\bibitem{Katori} H. A. Katori, K. Katsumata, O. Petracic, W. Kleemann. T. Kato, Ch. Binek, Phys. Rev. B
63 (2001) 132408.
\bibitem{Narumi} Y. Narumi, K. Katsumata, T. Nakumura, Y. Tanaka, S. Shimomura, T. Ishikawa, M.
Yabashi, J. Phys.: Condens. Matter 16 (2004) L57.
\bibitem{Engelstad} P. E. Engelstad and K. Yamada, Phys. Rev. B 52 (1995) 13029.
\bibitem {Durin} G. Durin, M. Bonaldi, M. Cerdonio, R. Tommasini, S. Vitale, J. Magn. Magn.
Mater. 101 (1991) 89.
\bibitem {Raap} M. B. F. van Raap, F.H. Sanchez, C.E.R. Torres, L. Casas, A. Roig, E. Molins, J. Phys.:
Condens. Matter 17 (2005) 6519.
\bibitem{Zukovic1} M. Zukovic, A. Bobak ve T. Idogaki, J. Magn. Magn. Mater. 188 (1998) 52.
\bibitem{Zukovic2} M. Zukovic, A. Bobak ve T. Idogaki, J. Magn. Magn. Mater. 192 (1999) 363.
\bibitem{Binek1} Ch. Binek, W. Kleemann, Phys. Rev. Lett. 72 (1994) 1287; Ch. Binek, W. Kleemann, Acta Phys. Slovaca 44 (1994) 435.
\bibitem{Binek2} Ch. Binek, M.M.P. de Azedevo, W. Kleemann, D. Bertrand, J. Magn. Magn. Mater. 140–144 (1995) 1555.

\end{thebibliography}
\end{document}